\documentclass[aip,jap,reprint,superscriptaddress
]{revtex4-1}
\usepackage{graphicx}
\usepackage{amsmath,amsfonts,amssymb}
\usepackage{xcolor}
\usepackage{epstopdf}
\renewcommand{\epsilon}{\varepsilon}

\draft 

\begin{document}


\title{Measuring complete band diagrams of non-ideal heterointerfaces by combining ellipsometry and photoemission spectroscopy}




\author{Andrea Crovetto}
\email[]{Electronic mail: ancro@fysik.dtu.dk}
\affiliation{DTU Nanotech, Technical University of Denmark, DK-2800 Kgs. Lyngby, Denmark}
\affiliation{SurfCat, Department of Physics, Technical University of Denmark, DK-2800 Kgs.~Lyngby, Denmark}

%

\begin{abstract}

In this work, we show that spectroscopic ellipsometry can be combined with photoemission spectroscopy to obtain complete interface band diagrams of non-ideal semiconductor heterointerfaces, such as interfaces between thin-film polycrystalline materials. The non-destructive ellipsometry measurement probes the near-interface band gap of the two semiconductors (including the buried semiconductor) after the interface has formed. This is important in the non-ideal case where chemical processes during interface growth modify the electronic properties of the two separated surfaces. Knowledge of near-interface band gaps improves accuracy in conduction band offset measurements of non-ideal interfaces, and it sheds light on their device physics. Both of those positive outcomes are demonstrated in the Cu$_2$ZnSnS$_4$/CdS interface used here as a case study, where the band gap of both materials decreases by up to 200~meV from the bulk to the near-interface region. This finding reveals a preferential electron-hole recombination channel near the interface, and it yields corrected values for the interfacial conduction band offset.

\end{abstract}

\pacs{}

\maketitle 


\section{Introduction}
Conduction band offset measurements of semiconductor heterointerfaces are typically based on a set of direct photoemission spectroscopy measurements.~\cite{Klein2015} Those measurements are sensitive to the energy position of the valence band maximum at the semiconductor surface, that is, within a depth of 1-20 nm depending on the excitation wavelength (UV light or x-rays). To access the conduction band positions, two possibilities exist. The first is simply to measure the bulk band gap of each semiconductor by an optical technique, such as UV-visible transmission/reflection spectroscopy.~\cite{Santoni2013} The second is to perform two complementary inverse photoemission spectroscopy measurements (IPES), which yield the energy position of the conduction band minima of the two semiconductor surfaces with an analysis depth of a few nm.~\cite{Bar2011a} In the first (second) case, the bulk (surface) band gaps of the two semiconductors are added to the valence band offset to determine the conduction band offset.

However, formation of a polycrystalline semiconductor heterointerface often results in considerable interdiffusion of chemical species across the interface. Various authors have reported significant interdiffusion at interfaces of interest for thin-film heterojunction solar cells, such as CdTe/CdS,~\cite{Enriquez2007} Cu(In,Ga)Se$_2$/CdS,~\cite{Heske1999} Cu$_2$ZnSn(S,Se)$_4$/CdS,~\cite{Baer2017} and Cu$_2$O/ZnO.~\cite{Wilson2014} Interdiffusion may be enhanced by heating the interface during or after growth, by energetic particle bombardment during growth, by the presence of grain boundaries as a preferential diffusion channel, and by the presence of voids and extended defects in the polycrystalline films.
Due to the diffusion and lattice incorporation of elements from the foreign semiconductor, the band structure of each semicondutor in the near-interface region may be significantly different from the respective bulk and near-surface band structure.

In this paper, we apply spectroscopic ellipsometry to non-destructively measure the near-interface band gaps of two heterojunction semiconductors \textit{after} interface formation, including the near-interface band gap of the buried semiconductor. Conduction band offsets can then be derived by adding the measured interface band gaps to the valence band offset determined with a standard photoemission-based method. Besides improving the accuracy of a conduction band offset measurement, knowledge of the interface band gaps of a semiconductor heterojunction can help understand its device physics. In fact, both band gap widening~\cite{Bar2008} and band gap narrowing~\cite{Crovetto2017} at the interface can have major consequences at the device level.

\section{Sample preparation}
Cu$_2$ZnSnS$_4$/CdS (CZTS/CdS) interfaces processed at different temperatures were employed as a case study for the proposed experimental method.
CZTS thin films (about 100 nm) were deposited on a soda lime glass substrate by pulsed laser deposition. The deposited films were further sulfurized at 550$^\circ$C for 10~min. Process details and film characterization are available elsewhere.~\cite{Cazzaniga2017} Three CdS thin films (20-40~nm) were then grown on nominally identical glass/CZTS substrates by chemical bath deposition at 55$^\circ$C, 75$^\circ$C, and 95$^\circ$C. Chemical bath deposition is the standard technique used in the highest efficiency CZTS solar cells.~\cite{Tajima2016a,Liu2017b}
After CdS deposition, each of the three samples was cut in two halves, and one half of each sample was post-annealed for 20 minutes at 300$^\circ$C in an Ar atmosphere.
A total of six samples was therefore available for the present study. As will be shown in the next section, simultaneous treatment of the data measured on several, slightly different samples (multi-sample analysis) can considerably improve the reliability of ellipsometry results.

\section{Measurement of near-interface band gaps}
\subsection{Theory}
In an ellipsometry measurement in reflection mode with incidence angle $\theta$, the polarization state of the incident beam can be described by the amplitude and phase of the two polarization components $s$ (perpendicular to the plane of incidence) and $p$ (contained in the plane of incidence). One can then define $r_\mathrm{s}(E)$ and $r_\mathrm{p}(E)$ as the complex reflection coefficients of the sample as a function of photon energy for the $s$ and the $p$ polarization components respectively.
By measuring the polarization state of the reflected beam and comparing it to the known polarization state of the incident beam, the ellipsometer determines the complex ratio $r_\mathrm{p}(E) / r_\mathrm{s}(E)$ of the two reflection coefficients. That ratio is typically expressed by the two real numbers $\Psi(E)$ and $\Delta(E)$, where

\begin{equation}
\frac{r_\mathrm{p}(E)} {r_\mathrm{s}(E)} = \tan[\Psi(E)]\, e^{i\Delta(E)}
\label{eq:ellipsometry}
\end{equation}

For a sample consisting of a number of stacked flat layers, each with a known thickness $d$ and dielectric function $\varepsilon(E)$, the overall reflection coefficients $r_\mathrm{p}(E)$ and $r_\mathrm{s}(E)$ at a given incidence angle $\theta$ can be determined by theory using the Fresnel equations to obtain the reflection coefficient of each interface and a correction factor to account for interference effects due to the finite thickness of each layer (Fig.~\ref{fig:method}(a)). Therefore, $\Psi(E)$ and $\Delta(E)$ spectra can be completely determined by theory as a function of the thicknesses and dielectric functions of all layers. If the problem is inverted, that is, if some of the thicknesses and dielectric functions involved are unknown, they can in principle be determined by regression analysis through least-squares fitting of the \textit{measured} $\Psi(E)$ and $\Delta(E)$ spectra using the unknown quantities as fitting parameters. This is the principle of spectroscopic ellipsometry.~\cite{Fujiwara2007}

\begin{figure}[t!]
\centering%
\includegraphics[width=\columnwidth]{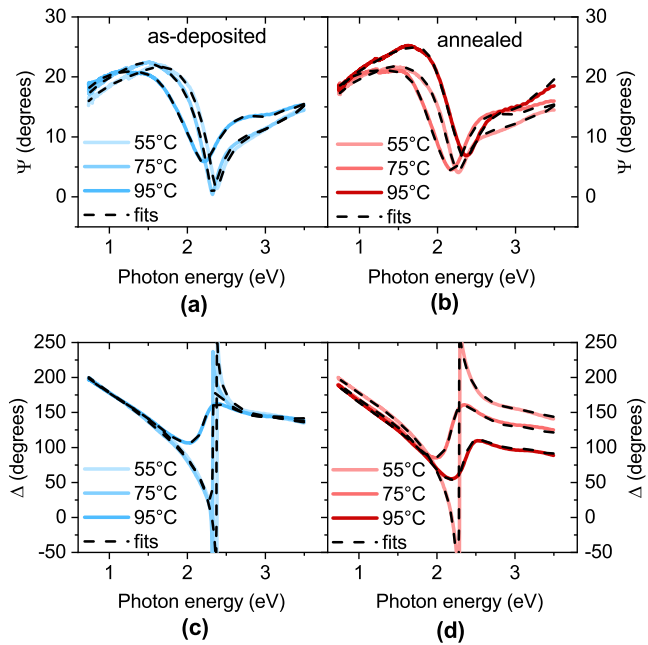}
\caption{$\Psi(E)$ spectra (a,b) and $\Delta(E)$ spectra (c,d) of the six investigated samples, of which three are measured as-deposited (a,c) and three are measured after post-anneling (b,d). For readability, only the spectra taken at an angle of incidence of 60$^\circ$ are included. The measured and fitted spectra at all (six) incidence angles are shown in the Supplementary Material.}
\label{fig:psi_delta}
\end{figure}

The goal of this work is to determine the band gaps of CZTS and CdS near the CZTS/CdS interface, where they are potentially modified by chemical processes with respect to the bulk. Based on the discussion above, this can be done by spectroscopic ellipsometry with the following procedure. (i) Acquiring $\Psi(E)$ and $\Delta(E)$ spectra from a glass/CZTS/CdS sample; (ii) modeling the CZTS and CdS layers as several sub-layers, each of which has an independent dielectric function (thus an independent band gap, Fig.~\ref{fig:method}(a)); (iii) parameterizing the dielectric function of each layer with a number of fitting parameters which, together with the unknown layer thickesses, are used to fit the measured $\Psi(E)$ and $\Delta(E)$ spectra (Fig.~\ref{fig:psi_delta}); (iv) extracting the band gap value of each layer based on the shape of the fitted imaginary part of the dielectric function near the absorption onset (Fig.~\ref{fig:method_onsets}(a)).

\subsection{Experimental details and analysis method}
$\Psi(E)$ and $\Delta(E)$ spectra were acquired in reflection mode with a rotating compensator ellipsometer (M-2000, J.A. Woollam Co.). The spot size was about 200~$\mu$m $\times$ 300~$\mu$m and the photon energy range was $0.78~\mathrm{eV} \leq E \leq 3.50~\mathrm{eV}$. The measurement was repeated at six different incidence angles $\theta$  for each sample ($45^\circ \leq \theta \leq 70^\circ$) to provide a large dataset for more robust fitting. The CompleteEase software package (version 5.06 - J.A. Woollam Co.) was used for fitting work.

\begin{figure}[t!]
\centering%
\includegraphics[width=\columnwidth]{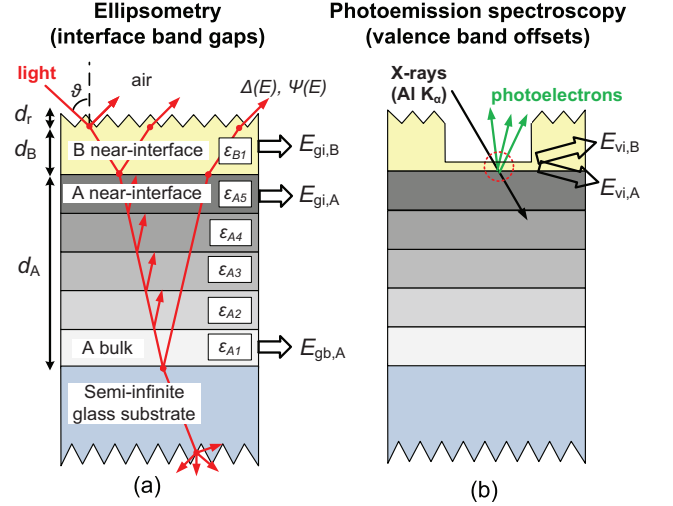}
\caption{(a): Scheme of the optical model used to extract the band gaps of CZTS ("A") and CdS ("B") by fitting the ellipsometry spectra $\Psi(E)$ and $\Delta(E)$. The band gaps $E_\mathrm{gb,A}$, $E_\mathrm{gi,A}$, and $E_\mathrm{gi,B}$ are derived from the fitted dielectric functions $\varepsilon_\mathrm{A1}$, $\varepsilon_\mathrm{A5}$, and $\varepsilon_\mathrm{B1}$ of the corresponding layers. (b): Scheme of the XPS measurement employed to determine the valence band offset $\mathrm{VBO} = E_\mathrm{vi,B}-E_\mathrm{vi,A}$. Refer to the main text for the meaning of the symbols.}
\label{fig:method}
\end{figure}

The layer model employed in this study is shown in Fig.~\ref{fig:method}(a). One important remark is that, while the expected $\Psi(E)$ and $\Delta(E)$ of a known sample are unique and can be calculated analytically, the uniqueness of the solution to the inverse problem (i.e., fitting the measured $\Psi(E)$ and $\Delta(E)$ to determine unknown properties of the samples) is not guaranteed. Hence, the sample should be modeled with the smallest possible number of fitting parameters, and the largest possible number of independent spectra should be fitted simultaneously against a consistent model. For this reason, the 36 $\Psi(E)$ spectra and the 36 $\Delta(E)$ spectra measured in this work (6 samples times 6 angles of incidence) were all fitted simultaneously, as detailed later in the article as well as in the Supplementary Material.

The CZTS layer ("A") has unknown thickness $d_\mathrm{A}$ (fitting parameter) and unknown, possibly depth-dependent dielectric function due to interdiffusion. To model a depth-dependent dielectric function at the smallest possible expense of fitting parameters, we slice the CZTS layer into five separate layers of equal thickness (Fig.~\ref{fig:method}(a)). The dielectric function of the bottom layer $\varepsilon_\mathrm{A1}(E)$ is parameterized with a Kramers-Kronig-consistent b-spline function with 3 nodes/eV, corresponding to 10 fitting parameters in the measured range $0.78~\mathrm{eV} \leq E \leq 3.50~\mathrm{eV}$, similarly to previous ellipsometry work on CZTS.~\cite{Crovetto2015,Crovetto2016} The possible band gap change from the bottom layer (bulk CZTS) to the top layer (near-interface CZTS) is modeled by allowing the b-spline node located at 1.68~eV to vary linearly from the bottom layer to the top layer. All other nodes are kept constant through the layers. The b-spline node at 1.68~eV in the dielectric function $\varepsilon_\mathrm{A5}(E)$ of the top CZTS layer is the only additional fitting parameter with respect to the case of a depth-independent dielectric function.

The dielectric function $\varepsilon_\mathrm{B1}(E)$ of the CdS layer ("B") could also be depth-dependent in principle. However, slicing CdS into sub-layers with varying dielectric function only led to marginal improvement of the mean squared error (MSE) of the fit with poor reproducibility of the best-fit parameters upon variation of the initial guess. This may indicate that the band gap of CdS is roughly depth-independent up to a few tens of nm away from the interface with CZTS, or that a depth-dependent dielectric function is difficult to capture for such a thin layer. CdS is thus modeled as a single layer with unknown thickness $d_\mathrm{B}$ and unknown dielectric function $\varepsilon_\mathrm{B1}(E)$. The imaginary part of $\varepsilon_\mathrm{B1}(E)$ is parameterized with one simplified Tauc-Lorentz oscillator and one pole in the ultraviolet, and the real part of $\varepsilon_\mathrm{B1}(E)$ is derived by subsequent Kramers-Kronig integration.
The parameters of a Tauc-Lorentz oscillator are the amplitude $A$, broadening $B$, and energy $E_\mathrm{o}$ of a Lorentzian function, plus a band gap energy parameter $E_\mathrm{g}$ and a high-frequency dielectric function parameter $\varepsilon_{\infty}$.~\cite{Jellison1996} Enforcing the conditions $\varepsilon_{\infty}=1$ and $E_\mathrm{o} = E_\mathrm{g}$ does not have noticeable effects on the MSE of our study. This limits the number of fitted parameters for the Tauc-Lorentz oscillator to 3. The additional node in the ultraviolet has amplitude $A_\mathrm{uv}$ and energy $E_\mathrm{uv}$ as fitting parameters.

Surface roughness has a strong influence on the $\Psi(E)$ and $\Delta(E)$ spectra and can complicate ellipsometry analysis if it exceeds a few nm.~\cite{Fujiwara2016} Roughness was minimized in the sample preparation phase by intentionally growing relatively thin (100~nm) CZTS layers. In the fitting phase, the roughness layer is modeled by means of Bruggeman's effective medium theory assuming a $50\%/50\%$ mix of air and CdS in the roughness layer.~\cite{Fujiwara2007} This requires 1 fitting parameter $d_\mathrm{r}$ for the thickness of the roughness layer (Fig.~\ref{fig:method}(a)). The dielectric function of the glass substrate was determined from a separate ellipsometry measurement on bare glass and is kept fixed when fitting CZTS/CdS ellipsometry spectra. The thickness of the glass substrate does not influence the $\Psi(E)$ and $\Delta(E)$ spectra, because opaque tape was applied to the back side of the glass to suppress back-side reflection.~\cite{Synowicki2008}

To further reduce the chance of overparameterizing the fitting problem, the $\Psi(E)$ and $\Delta(E)$ spectra from the six samples investigated in this work are all fitted simultaneously, using an approach known as multi-sample analysis.~\cite{Hilfiker2008,Fujiwara2007} The parameters that are expected to vary significantly from sample to sample are fitted independently in each sample. Those 6 parameters are $d_\mathrm{A}$, $d_\mathrm{B}$, $d_\mathrm{r}$, $E_\mathrm{o}$, as well as the nodes of $\varepsilon_\mathrm{A1}(E)$ and $\varepsilon_\mathrm{A5}(E)$ at 1.68~eV. Conversely, the parameters that are expected to be roughly constant from sample to sample are forced to have the same best-fit value across samples. Those 13 parameters are $A$, $B$, $A_\mathrm{uv}$, $E_\mathrm{uv}$, and the 9 remaining nodes of $\varepsilon_\mathrm{A1}(E)$.

\subsection{Results}
\begin{figure}[t!]
\centering%
\includegraphics[width=\columnwidth]{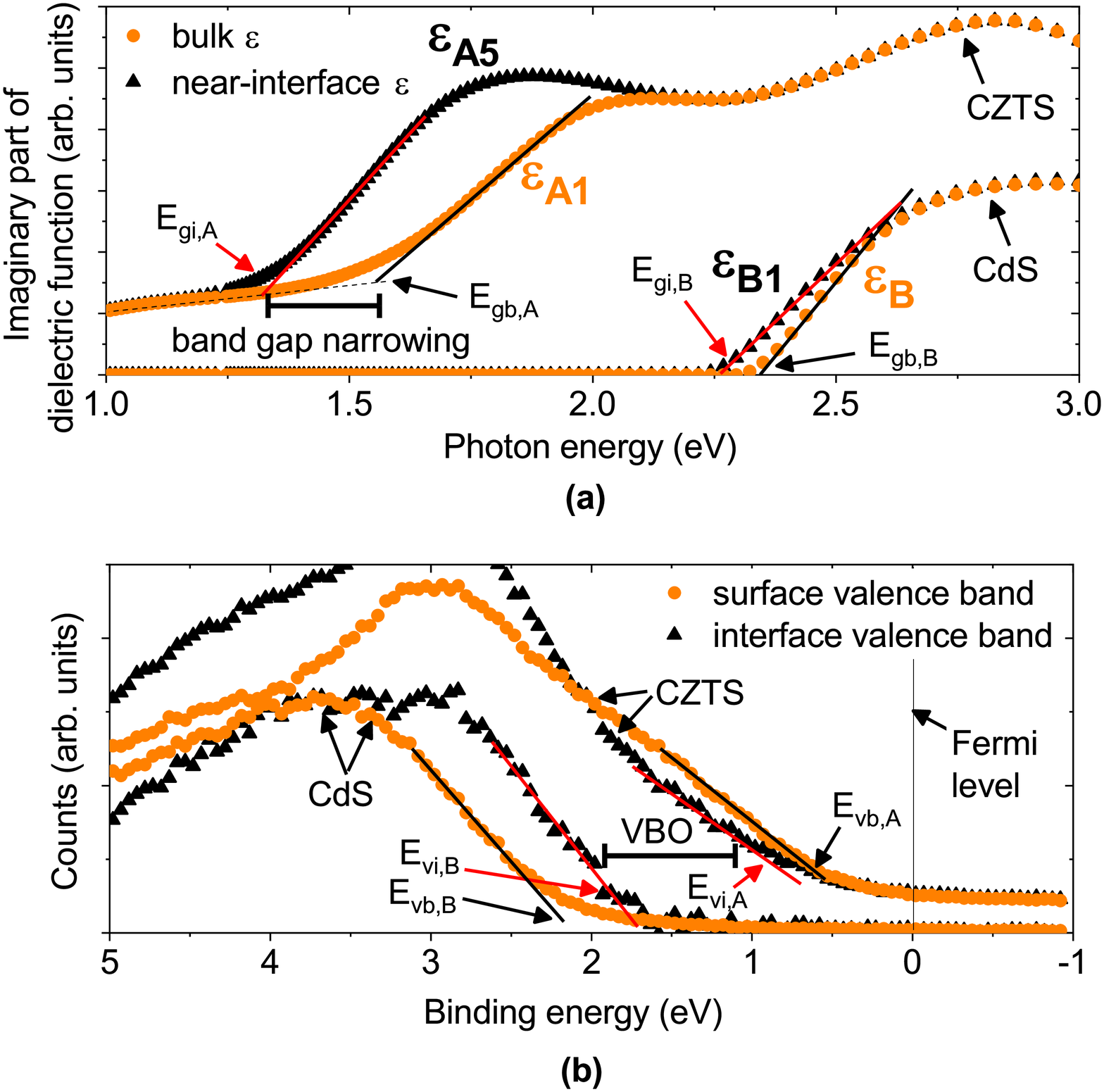}
\caption{(a): The fitted imaginary part of the dielectric function of various layers of one sample in this study (CdS deposition temperature of 55$^\circ$C, with post-annealing). Note that $\varepsilon_\mathrm{B}$ is obtained from a separate ellipsometry measurement of a CdS film grown on ITO-coated glass. The corresponding band gaps are extracted by extrapolating the absorption onset with a straight line. Refer to Fig.~\ref{fig:method}(a) and to the main text for interpreting the symbols. (b) Relevant photoemission onsets for the same sample as above. Valence band positions are obtained by extrapolating photoemission onsets with a straight line. Note that $E_\mathrm{vi,B}$ is derived from the same photoemission spectrum as $E_\mathrm{vi,A}$ by the valence band difference method.~\cite{Klein1997} Note also that the two bare-surface valence band spectra are measured on an in-situ-cleaned bare CZTS surface and on in-situ-cleaned bare CdS surface on an ITO-coated glass substrate. Refer to Fig.~\ref{fig:method}(b) and to the main text for interpreting the symbols.}
\label{fig:method_onsets}
\end{figure}

While the multi-sample analysis approach yields a somewhat higher MSE (11.0) compared to separate fitting of each sample, it also results in a dramatic decrease of the error bars of the best-fit parameters, which are defined based on 90\% confidence intervals. As detailed in the Supplementary Material, the error bars of the layer thicknesses are roughly $\pm 1\%$; the error bars of the $E_\mathrm{o}$ parameter, related to the CdS band gap, are roughly $\pm 0.1\%$; the error bars of the $\varepsilon_\mathrm{A1}(1.68)$ node, related to the CZTS bulk band gap, are roughly $\pm 1\%$; and the error bars of the $\varepsilon_\mathrm{A1}(1.68)-\varepsilon_\mathrm{A5}(1.68)$ quantity, related to the CZTS near-interface band gap, are between $\pm 1\%$ and $\pm 10\%$.  The best-fit values of all fitting parameters are shown in the Supplementary Material. In particular, knowledge of the best-fit values of the dielectric function parameters for the two materials allows derivation of their band gaps. Band tailing in CZTS absorbers is a well-known issue, which hinders accurate determination of the band gap by standard Tauc plots.~\cite{Siebentritt2015} Band tails are clearly visible in Fig.~\ref{fig:method_onsets}(a) as the imaginary part of the CZTS dielectric function does not go to zero below the expected band gap energy. Therefore, we simply estimate the bulk- and near-interface band gap of CZTS ($E_\mathrm{gb,A}$ and $E_\mathrm{gi,A}$ respectively) by extrapolating the imaginary part of the corresponding dielectric function ($\varepsilon_\mathrm{A1}$ and $\varepsilon_\mathrm{A5}$ respectively) with a straight line, as shown in Fig.~\ref{fig:method_onsets}(a). For consistency, the same procedure is applied to determine the bulk band gap of CdS ($E_\mathrm{gb,B}$) from $\varepsilon_\mathrm{B}$, and the near-interface band gap of CdS ($E_\mathrm{gi,B}$) from $\varepsilon_\mathrm{B1}$. This simple approach has two advantages: (i) it yields bulk band gaps in good agreement with the CZTS band gap as determined by quantum efficiency analysis of a complete solar cell (1.56~eV versus 1.53~eV);~\cite{Cazzaniga2017} (ii) it removes part of the ambiguity in the extrapolation of the dielectric function near the absorption onset, since the band tail is easily distinguished from the band gap (Fig.~\ref{fig:method_onsets}(a)). Note that the quantity $\varepsilon_\mathrm{B}$, and therefore the bulk band gap of CdS, is obtained from a separate ellipsometry measurement of a CdS film grown on ITO-coated glass.

\begin{figure}[t!]
\centering%
\includegraphics[width=\columnwidth]{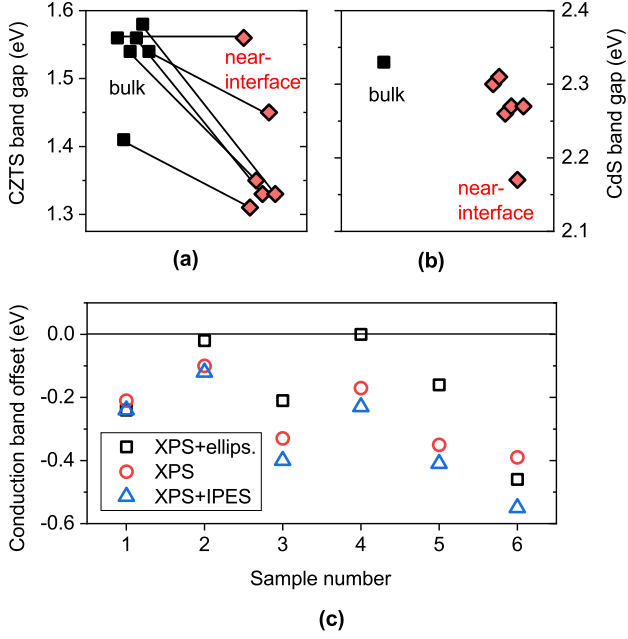}
\caption{(a) The bulk- and near-interface band gaps of CZTS as determined by ellipsometry for the six samples investigated in this study. The solid lines connect the bulk band gap ($E_\mathrm{gb,A}$) and the near interface band gap ($E_\mathrm{gi,A}$) of the same sample. (b) The bulk- and near-interface band gaps of CdS as determined by ellipsometry for the same samples. (c) Conduction band offset of each sample calculated with Eq.~\ref{eq:XPS+ell} (squares), with Eq.~\ref{eq:XPS} (circles), and with Eq.~\ref{eq:XPS+IPES} (triangles).}
\label{fig:errors}
\end{figure}

A shown in Figs.~\ref{fig:errors}(a),~\ref{fig:errors}(b), the general trend for both materials is band gap narrowing in the near-interface region. Especially the band gap of CZTS decreases by several hundred meV from bulk to the interface. This effect alone can have serious consequences at the device level.~\cite{Crovetto2017}

\section{Measurement of valence band offsets}
To determine the conduction band offsets of the six samples using the near-interface band gaps measured by ellipsometry, valence band offsets must also be measured. An established experimental technique based on x-ray photoelectron spectroscopy (XPS) was employed,~\cite{Klein1997} using a Thermo Scientific K-Alpha instrument with a monochromatized Al K$_\alpha$ x-ray source and a spot size of roughly 400~$\mu$m. The binding energy scale was calibrated with the Fermi level of an in-situ-cleaned Au sample, and confirmed by the position of the adventitious C~1s peak on as-prepared samples.
The different CZTS/CdS samples were progressively etched with low-energy Ar$^+$ ions (200~eV) and photoemission spectra of the valence band region
were recorded after each etching step (Fig.~\ref{fig:method}(b)). This ion energy has been shown not to alter the valence band features of CdS and CZTS.~\cite{Santoni2013} Valence band positions with respect to the Fermi level (zero binding energy) were measured by extrapolating the energy of the photoemission onset of each material with a straight line~\cite{Bar2011a,Crovetto2016g} as shown in Fig.~\ref{fig:method_onsets}(b). The first spectrum with a recognizable photoemission onset of CZTS was used to determine the interface valence band position of CZTS with respect to the Fermi level ($E_\mathrm{vi,A}$). The interface valence band position of CdS ($E_\mathrm{vi,B}$) was derived from the same spectrum by the technique of valence band difference spectra~\cite{Klein1997} or "direct VBO method".~\cite{Crovetto2017b} The interface valence band offset (VBO) was simply calculated as VBO = $E_\mathrm{vi,B} - E_\mathrm{vi,A}$ (Fig.~\ref{fig:method_onsets}(b)). The valence band difference method has been applied to CZTS/CdS interfaces before~\cite{Santoni2013} and it involves deconvoluting the CZTS valence band signal from the total photoemission signal in order to determine the energy of the superimposed CdS photoemission onset.~\cite{Klein1997} The valence band positions of a bare CZTS surface ($E_\mathrm{vb,A}$) and of a bare CdS surface ($E_\mathrm{vb,B}$) were also measured for reference and are shown in Fig.~\ref{fig:method_onsets}(b). They were measured by XPS on a bare CZTS film on glass and on a CdS film deposited on ITO-coated glass respectively, after removal of the native oxide layer by Ar$^+$ etching at 200~eV.

\section{Discussion and comparison with other methods}
The combination of ellipsometry- and XPS measurements makes it possible to determine conduction band offsets from valence band offsets using the relevant band gaps for the problem, i.e., the near-interface band gaps. As shown in Fig.~\ref{fig:errors}(c), the measured conduction band offsets (CBO) across the six samples span over a 0.4~eV range, partially due to differences in their near-interface band gaps. It can be interesting to compare the CBO values determined by the present method to the values that would be obtained by more established methods. For the present method (XPS+ellipsometry):
\begin{equation}
\mathrm{CBO} = \mathrm{VBO} + E_\mathrm{gi,B} - E_\mathrm{gi,A}
\label{eq:XPS+ell}
\end{equation}
\\
For a method based on the XPS-determined valence band offset and bulk band gaps:
\begin{equation}
\mathrm{CBO} = \mathrm{VBO} + E_\mathrm{gb,B} - E_\mathrm{gb,A}
\label{eq:XPS}
\end{equation}
\\
For a method based on the XPS-determined valence band offset and the IPES-determined surface band gaps, assuming that the surface band gap measured on a bare CZTS surface is equal to $E_\mathrm{gb,A}$, and that the surface band gap measured on a thin CdS layer deposited on CZTS is equal to $E_\mathrm{gi,B}$:~\cite{Bar2011a}
\begin{equation}
\mathrm{CBO} = \mathrm{VBO} + E_\mathrm{gi,B} - E_\mathrm{gb,A}
\label{eq:XPS+IPES}
\end{equation}

Fig.~\ref{fig:errors}(c) compares the CBOs determined by Eqs.~\ref{eq:XPS+ell}-\ref{eq:XPS+IPES}. The underestimation of the CBO by the XPS+IPES method (Eq.~\ref{eq:XPS+IPES}) can be explained by the fact that the XPS+IPES method caputures near-interface band gap narrowing in CdS, but not in CZTS.~\cite{Bar2011a} The less severe CBO underestimation by the XPS method with separately determined bulk band gaps (Eq.~\ref{eq:XPS}) can be explained by error cancellation, since band gap changes of both CZTS and CdS near the interface are in the same direction (i.e., towards band gap narrowing).

\begin{figure}[t!]
\centering%
\includegraphics[width=\columnwidth]{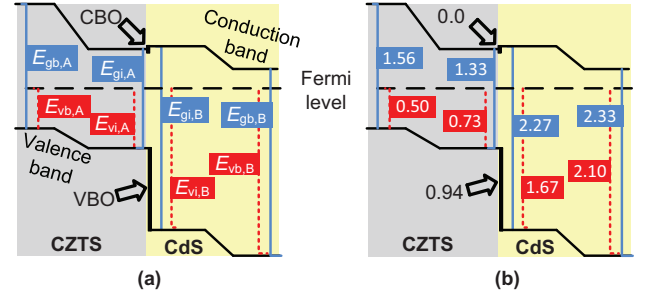}
\caption{(a) Key for relating the quantities measured in this study to interface band diagrams. (b) The specific interface band diagram extracted for the sample with a CdS deposition temperature of 55$^\circ$ and post-anneling.}
\label{fig:band_diagrams_method}
\end{figure}

Finally, detailed interface band diagrams can be drawn based on the quantities measured in this work, as illustrated in Fig.~\ref{fig:band_diagrams_method}(a). Taking the interface band diagram of the sample with a CdS deposition temperature of 55$^\circ$C and post-anneling as an example (Fig.~\ref{fig:band_diagrams_method}(b)), interesting features that influence the physics of the resulting device are found. The flat conduction band offset is a beneficial feature for electron transport across the interface and minimization of interface recombination.~\cite{Crovetto2017b} The downwards band bending from CZTS to CdS is also a beneficial feature, as it drives electrons and holes towards their respective contacts. Band gap narrowing in CZTS is, however, a detrimental feature, as it decreases the barrier to recombination in the near-interface region, thus potentially decreasing the maximum achievable solar cell open-circuit voltage.~\cite{Crovetto2017} Understanding the specific chemical mechanisms responsible for band gap narrowing and their dependence on process conditions is not straightforward and will thus require a separate study. Here we simply note that near-interface band gap narrowing of CZTS and CdS is observed consistently in all samples, regardless of their different processing conditions.

\section{Method limitations}
Our proposed experimental method certainly has a number of limitations. First of all, any effect that modifies the band gaps of the materials only in a very shallow region of a few nm across the interface is very difficult to detect by ellipsometry, since the effect of dielectric function changes on $\Psi(E)$ and $\Delta(E)$ spectra for very thin layers can be indistinguishable from thickness changes.~\cite{Archer1962} Hence, the present method is only sensitive to "near-interface" band gap changes (tens of nm), which can be related, e.g., to relatively deep interdiffusion or to phase segregation after post-annealing. This is the reason for the expression "near-interface band gap" used throughout this article. Since ellipsometry is a model-based indirect technique, the quality of the ellipsometry results strongly depends on the correctness of the layer model and on careful selection of fitting parameters, which is a trade-off between an oversimplified and an overparametrized model.~\cite{Fujiwara2007} Finally, the sample preparation phase is critical. Polycrystalline materials with relatively large, micron-sized grains can result in unacceptably high surface roughness, which renders ellipsometry analysis impossible.~\cite{Fujiwara2007} This issue can be circumvented by depositing thin layers of materials or by polishing the layers after deposition.

\vspace{0.5cm}

\section{Conclusion}
We have proposed an extension to a standard photoemission-based method for obtaining detailed interface band diagrams of semiconductor heterojunctions where chemical processes during interface growth or post-annealing modify the interface electronic properties with respect to the bulk or to the bare surface.
For the particular heterointerface selected to demonstrate the method (Cu$_2$ZnSnS$_4$/CdS), a band gap decrease of several hundred meV from bulk to interface was observed on both sides of the interface. The resulting device physics is significantly influenced by the band gap narrowing phenomenon. Adding the near-interface band gaps to the valence band offsets measured by photoemission spectroscopy significantly improves the accuracy of a standard conduction band offset measurement.


\section{Supplementary material}
See the Supplementary material for more details about the optical model used to fit ellipsometry spectra, for the best-fit values of all fitting parameters, and for plots of the experimental versus modeled ellipsometry spectra of all the investigated samples.

\section{Acknowledgements}
This work was supported by the Danish Council for Strategic Research, the Innovation Fund Denmark (File No. 5016-00102), and VILLUM Fonden (grant 9455).
We are grateful to Rebecca Ettlinger and Andrea Cazzaniga for assistance with sample preparation, as well as to Ole Hansen and J\o rgen Schou for useful discussions.

\end{document}